\documentclass[conference]{IEEEtran}
\IEEEoverridecommandlockouts
\usepackage{cite}
\usepackage{tabularx}
\usepackage{adjustbox}
 \usepackage{graphicx}
\usepackage{multirow}
\usepackage{textcomp}
\usepackage{colortbl}
\usepackage{hhline}
\usepackage{rotating}
\usepackage[hidelinks,colorlinks=true,citecolor=blue]{hyperref}
\usepackage[table,xcdraw,dvipsnames]{xcolor}
\usepackage{tabulary}
\usepackage{amsmath,amssymb,amsfonts,amsthm}
\usepackage{algorithmic}
\usepackage{array}

\allowdisplaybreaks
\usepackage{tikz}
\usepackage{tikz-cd}
\usepackage{stackengine}
\usepackage{scalerel}
\usepackage{fp}
\def\clipeq{\!\mathrm{=}\!}
\def\Lla{\Longleftarrow}
\def\Lra{\Longrightarrow}
\newcount\argwidthfackn
\newcount\clipeqwidth
\savestack\tempstack{\stackon{$\clipeq$}{}}%
\clipeqwidth=\wd\tempstackcontent\relax
\newcommand\xleftrightarrows[2]{%
\savestack\tempstack{\stackon{$\scriptstyle#1$}{$\scriptstyle#1$}}%
\argwidth=\wd\tempstackcontent\relax%
\FPdiv\scalefactor{\the\argwidth}{\the\clipeqwidth}%
  \FPsub\scalefactor{\scalefactor}{.4}
  \FPmax\scalefactor{\scalefactor}{.04}%
  \mathrel{%
  \stackunder[1pt]{\stackon[3pt]{$\Lla\hstretch{\scalefactor}{\clipeq}\Lra$}%
     {$\scriptstyle#1$}}{$\scriptstyle#2$}%
  }%
}
\usepackage{float}
\usepackage{placeins}
\usepackage{enumitem}
\usepackage{changepage}
\theoremstyle{definition}

\def\BibTeX{{\rm B\kern-.05em{\sc i\kern-.025em b}\kern-.08em
    T\kern-.1667em\lower.7ex\hbox{E}\kern-.125emX}}

\begin{document}
\title{Does Twinning Vehicular Networks Enhance Their Performance in Dense Areas?}

\author{
	\IEEEauthorblockN{Sarah Al-Shareeda\IEEEauthorrefmark{1}, Sema F. Oktug\IEEEauthorrefmark{1}, Yusuf Yaslan\IEEEauthorrefmark{1}, Gokhan Yurdakul\IEEEauthorrefmark{2}, and Berk Canberk\IEEEauthorrefmark{3}}
	\IEEEauthorblockA{\IEEEauthorrefmark{1}Faculty of Computer Engineering and Informatics, Istanbul Technical University, Turkey}
 \IEEEauthorblockA{\IEEEauthorrefmark{2}BTS Group, Turkey}
 \IEEEauthorblockA{\IEEEauthorrefmark{3}School of Computing, Engineering and The Built Environment, Edinburgh Napier University, UK}
	\{alshareeda, oktug, yyaslan\}@itu.edu.tr, yurdakulg@btsgrp.com, b.canberk@napier.ac.uk}
\maketitle
\begin{abstract}
This paper investigates the potential of Digital Twins (DTs) to enhance network performance in densely populated urban areas, specifically focusing on vehicular networks. The study comprises two phases. In Phase I, we utilize traffic data and AI clustering to identify critical locations, particularly in crowded urban areas with high accident rates. In Phase II, we evaluate the advantages of twinning vehicular networks through three deployment scenarios: edge-based twin, cloud-based twin, and hybrid-based twin. Our analysis demonstrates that twinning significantly reduces network delays, with virtual twins outperforming physical networks. Virtual twins maintain low delays even with increased vehicle density, such as 15.05 seconds for 300 vehicles. Moreover, they exhibit faster computational speeds, with cloud-based twins being 1.7 times faster than edge twins in certain scenarios. These findings provide insights for efficient vehicular communication and underscore the potential of virtual twins in enhancing vehicular networks in crowded areas while emphasizing the importance of considering real-world factors when making deployment decisions.
\end{abstract}

\section{Introduction and Contribution}\label{intro}
Intelligent Transportation Systems (ITS) have witnessed remarkable progress, revolutionizing urban mobility and optimizing transportation networks \cite{gong2023edge}. On the one hand, equipping vehicles with GPS units made it easy to track the travel history of vehicles, i.e., their recorded geospatial-temporal data; hence, it becomes easy to define the traffic flow in a particular region during a specific period. Several features can be predicted based on such trajectories' history, mainly identifying the region's most visited places or places witnessing long staying times by vehicles, also known as Points/Places of Interest (POIs), and identifying valuable insights into traffic patterns and vehicle behavior within urban environments. Numerous research efforts have been dedicated to estimating POIs, accelerating their extraction using Artificial Intelligence (AI) schemes, and utilizing the found POIs to enhance the travel experience of vehicles and address various transportation-related challenges \cite{massimo2020next}. Often, these POIs are dense and transportation-crowded locations, leading to higher accident rates. Hence, adding vehicle communication facilities, aka Vehicular Networks (VANETs) ecosystem, can improve transportation safety. However, the traditional network performance may score poorly in crowded areas because of the high witnessed contention for the communication medium. Many solutions have been presented in the literature to address the raised problems \cite{al2017enhancing}.

In recent years, creating a virtual ITS twin has been explored to accelerate and enhance their sought benefits. There have also been some efforts that address the creation of Digital Twins (DTs) for the vehicular communication aspect of ITS to improve the performance of VANETs \cite{yuan2022digital}. Despite such technological advancements, an unexplored area remains at the intersection of creating DTs for VANETs at crowded POIs. Here comes our contribution that addresses the integration of the DT for VANET to offset the physical network load to cyberspace, where smooth data-driven real-time decision-making and predictions of the system status can be made more accessible. Emulating vehicles' interactions at busy POIs makes optimizing traffic management possible, reducing congestion and improving transportation efficiency. In this regard, the main contributions of our study are:
\begin{itemize}
\item employing traffic flow data and AI-based schemes to identify and extract crowded POIs in Bursa, Turkey.
\item proposing the concept of vehicular network twins with diverse deployment strategies to enhance VANETs' performance in crowded areas.
\item conducting a case study of one extracted POI, providing evidence of the benefits of twinning vehicular networks.
\item evaluating the latency and computation speed performance indicators through SUMO and OMNeT++ simulations to offer quantitative insights into twin deployments' effectiveness.
\end{itemize}

Our research presents innovative solutions for managing communication at crowded POIs, shaping efficient transportation. Section \ref{background} reviews the literature on POIs extraction and creating VANETs virtual twins. Section \ref{design} describes the system model and its two phases. Section \ref{results} showcases the simulation case study and results discussion. Finally, Section \ref{conc} concludes with future extension ideas.

\section{Background and Literature Review}\label{background}
This section comprehensively surveys the literature on extracting POIs from geospatial data and creating DTs for VANETs. These domains have garnered significant research attention due to their wide-ranging transportation and urban planning applications. By examining the existing body of knowledge in these realms, we aim to gain insights into these areas that pave the easy construction of our planned scheme.

\subsection{POIs Extraction Background}
POIs can encompass well-known and frequently visited locations, including crowded hotspots where communication and transportation congestion are prevalent. Identification of POIs often relies on social networking and website/app popularity indices, and recommendation systems continuously update their POI databases based on these criteria \cite{ngamsa2020POInt,noorian2023bert,sanchez2022POInt}. On the other hand, as GPS data traces collected from vehicles have been explored to predict traffic congestion and enhance transportation efficiency, the same approach has been used to extract POIs. This approach utilizes the traffic flow history of a region to extract frequently visited GPS locations and POIs where vehicles spend considerable time, stop, or move slowly. By applying reverse geocoding, corresponding POIs are accurately defined, leading to improved recommendations. These collective efforts enhance the travel experience for users and vehicles, revealing potential synergies between traffic congestion prediction and POI identification methods. Several works have focused on extracting meaningful information from GPS trajectories to identify POIs and provide personalized recommendations. Palma et al. \cite{palma2008clustering} use speed-based analysis to identify trajectory stops and moves, revealing unexpected POIs. Zheng et al. \cite{zheng2009mining} mine interesting locations and travel sequences using multiple users' trajectories, considering factors like user visits and location interests. Rocha et al. \cite{rocha2010db} emphasize direction variation to identify interesting places. Huang and Gartner \cite{huang2010activity} utilize collaborative filtering for POI recommendations, incorporating visit frequencies and spatiotemporal behavior. Liu and Seah \cite{liu2015POInts} propose a density-based clustering algorithm considering popularity, temporal, and geographical features. Fiori et al. \cite{fiori2016decoclu} mine public transport networks to define routes using consensus density-based clustering. Keles et al. \cite{keles2017extracting} address POI identification challenges using Bayesian networks and various factors. Sun et al. \cite{sun2020city} propose an adaptive clustering framework considering varying POI granularities in different cities. Lastly, Massimo and Ricci \cite{massimo2023combining} customize recommendations for relevant following POIs based on similar visit behavior clusters.

These works collectively contribute to advancing POI extraction techniques, offering personalized and optimized user recommendations based on their vehicles' historical travel trajectories.

\subsection{Vehicular Networks and Creating Their DTs}\label{vanettwin}
VANETs are critical in enabling seamless communication within ITS. This allows vehicles and infrastructures to share real-time status and stay aware of their surroundings through Vehicle to Everything (V2X) communication. However, establishing reliable and efficient connections is challenging, especially with the frequent changes in physical network topology and connectivity disruptions. Addressing scalability hurdles caused by numerous vehicles generating real-time data adds complexity. Extensive research has been conducted to enhance the performance of VANETs and address these issues \cite{al2017enhancing}. Recently, the concept of DTs in ITS has emerged as a real-time emulation version of the system that can optimize and predict future actions. This emerging technology has led to the exploration of virtual twin representations of VANETs. By leveraging DTs, VANETs can utilize AI and advanced data analysis techniques to facilitate intelligent decision-making processes to optimize VANET operations. The comprehensive compilation of contributions to this concept reveals that current efforts focus on building and deploying twins and providing use cases of virtual VANETs to enhance physical networks. The use cases explored include enhancing autonomous driving tests \cite{han2022road}, improving lane changing and ramp merging scenarios \cite{fan2021digital}, addressing dynamics of vehicle grouping \cite{al2022vehicular}, and enhancing safety and warning decisions \cite{wang2023towards}. Additionally, some efforts have been presented to optimize communication performance, predict time delays, handle data distribution, enhance computing services, and improve content delivery efficiency in vehicular networks using their DTs \cite{sarah1, li2022digital}. Furthermore, recently, the security and privacy aspects of virtual twins of VANETs have been drawing the research attention to withstand the added attack surfaces of vehicular twins, \cite{he2022security}.

As evident from the presented research, significant efforts are directed toward constructing and implementing virtual vehicular network twins, investigating various use cases that leverage virtual VANETs to strengthen physical networks and enhance their various aspects.

\section{Our System Model for Twinning VANETs at Extracted Crowded POIs}\label{design}
\begin{figure*}[ht]
    \centering
    \includegraphics[width=.7\linewidth]{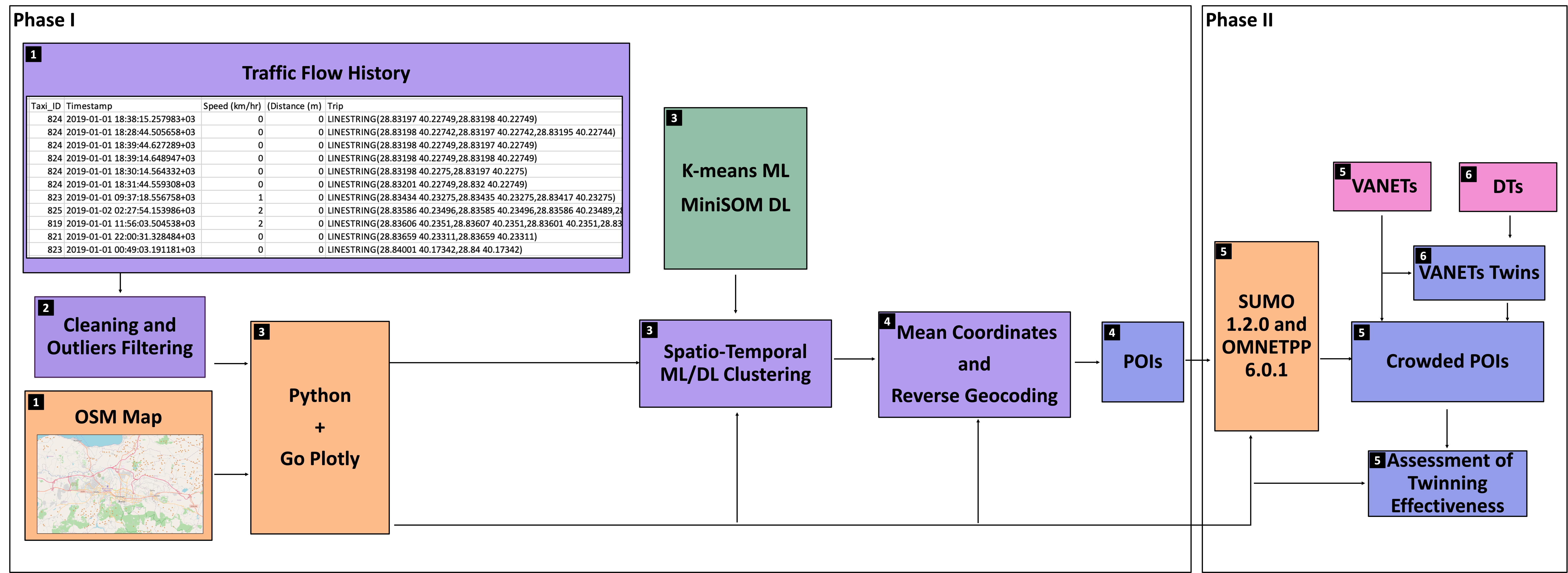}
    \caption{Detailed Description of Presented System Model}\label{fig:plan2}
\end{figure*}

This section introduces our novel system, specifically designed for deploying VANET twins at crowded POIs. Our addressed region is the busy city of Bursa, Turkey, within a geographical area spanning approximately 49.24km $\times$ 20.14km, bounded by ``28.456847, 40.103140, 29.388351, 40.318912" coordinates. We are leveraging geospatial GPS data collected from 221 taxis driving along the first two days of each month throughout 2019. The GPS data has recorded taxi IDs, latitude and longitude coordinates, speed, distance traveled, and spent time at these locations at 1,048,576 timestamps per month. Our proposed system, depicted in Fig. \ref{fig:plan2}, comprises two key phases: firstly, the extraction of crowded POIs through state-of-the-art spatiotemporal Machine Learning (ML) and Deep Learning (DL) clustering techniques, and secondly, envisioning the creation of twins for VANETs communication at these crowded extracted POIs. By harnessing this rich dataset and implementing our novel approach, we aim to revolutionize VANET communication, optimizing urban transportation for enhanced efficiency and safety.

\subsection{Phase I: AI-based Extraction of Crowded POIs}
\begin{figure}[!htbp]
    \centering
    \includegraphics[width=.6\linewidth]{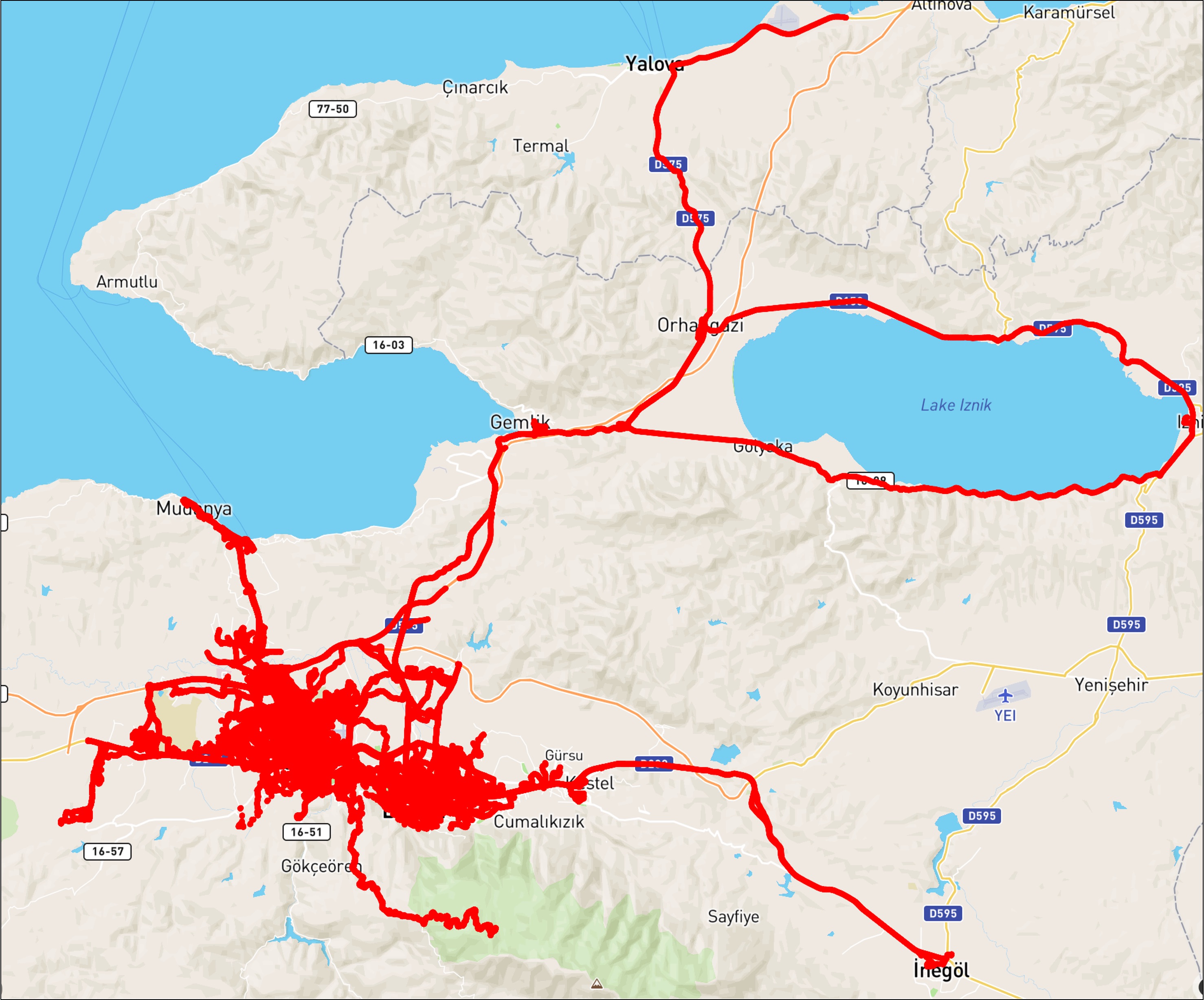}
    \caption{Pre-processing Initial Trajectories}\label{fig:pois1}
\end{figure}

\begin{figure}[!htbp]
    \centering
    \includegraphics[width=.6\linewidth]{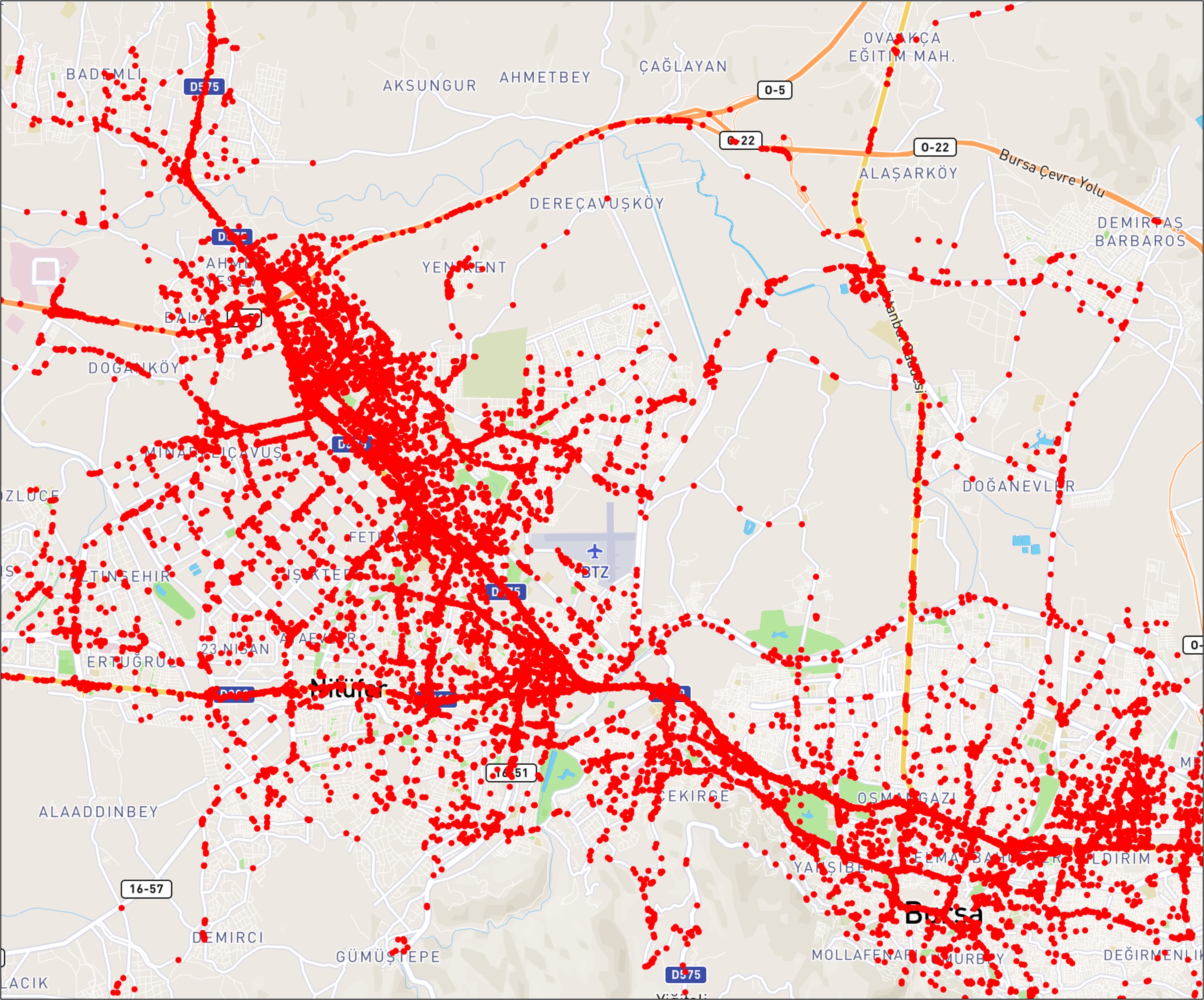}
    \caption{Post-processing Trajectories}\label{fig:pois2}
\end{figure}

\begin{figure}[!htbp]
    \centering
    \includegraphics[width=.6\linewidth]{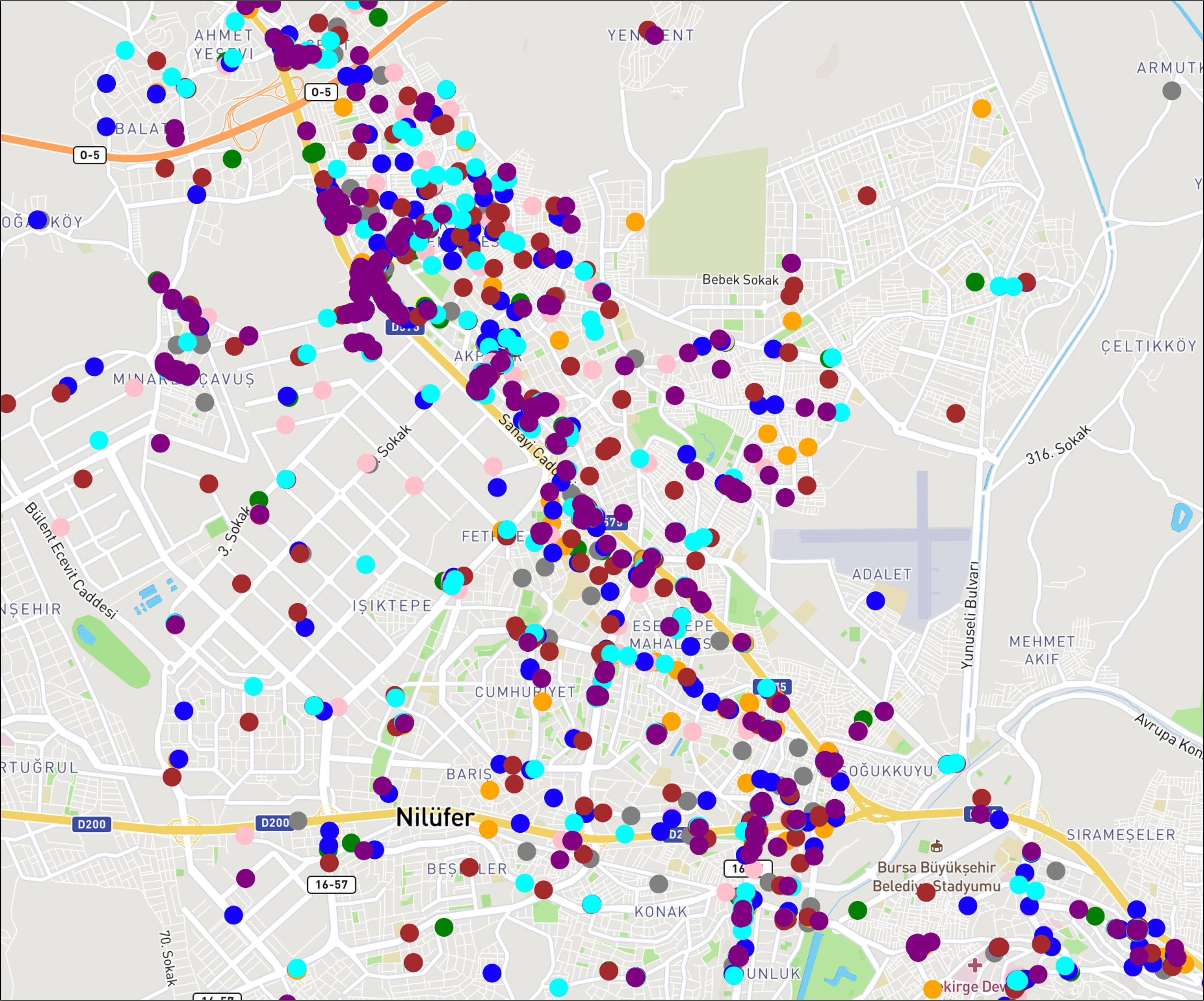}
    \caption{Clustered Trajectories}\label{fig:pois3}
\end{figure}

\begin{figure}[!htbp]
    \centering
    \includegraphics[width=.6\linewidth]{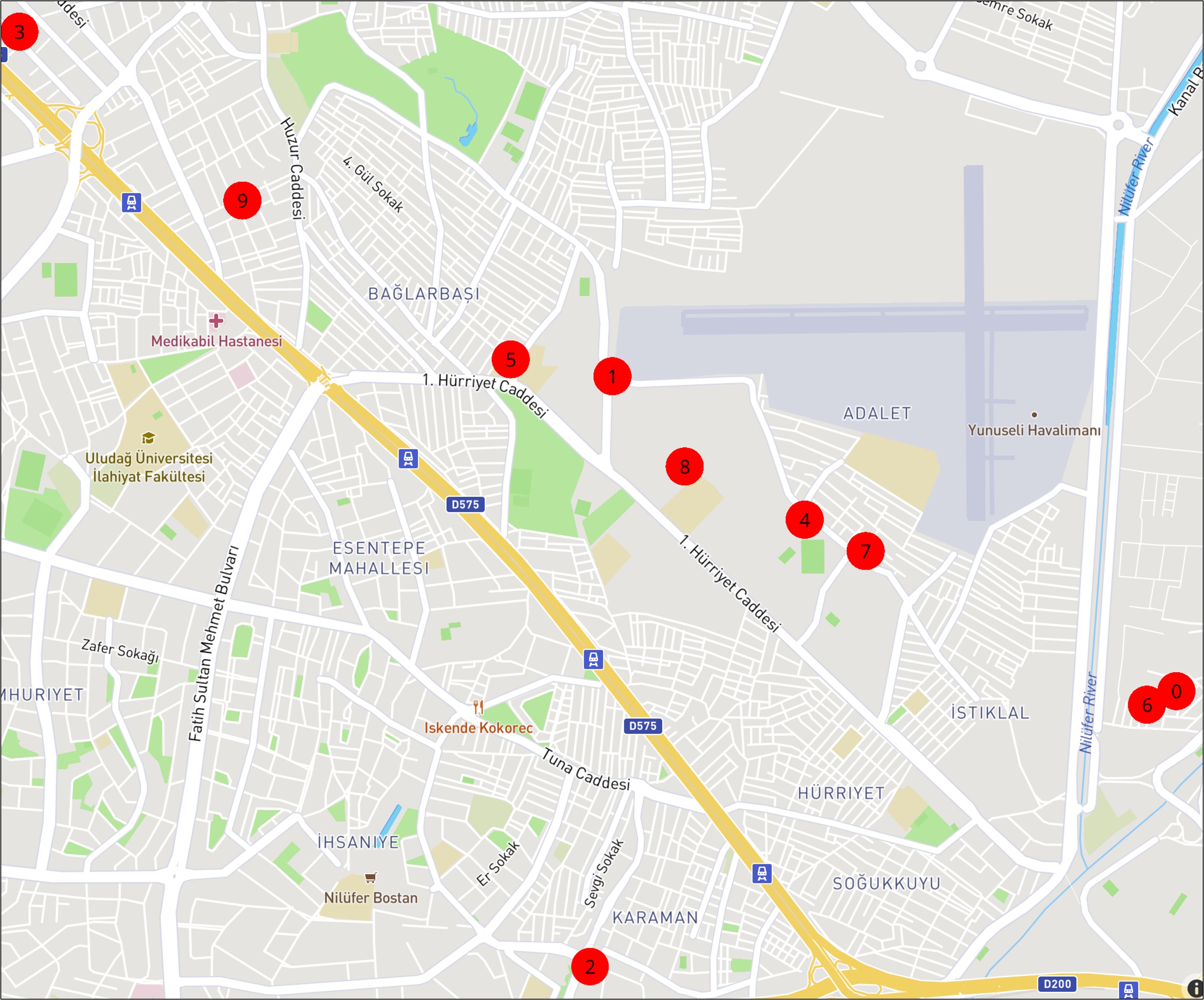}
    \caption{Extracted POIs}\label{fig:pois5}
\end{figure}

The geospatial taxi trajectories dataset of Bursa, Turkey, used in our study is extensive. However, we ensure data quality by filtering out outlier trips that deviate significantly from typical flow trajectories. We also limited our case study to January data to avoid extensive computation power requirements. After cleaning the dataset and preparing the simulation with the obtained OpenStreetMap map \cite{openstreetmap}, the next step involves selecting a spatiotemporal trajectory clustering approach \cite{ansari2020spatiotemporal}. These clustering methods group similar trajectories based on spatial and temporal characteristics, enabling the uncovering of patterns, anomaly detection, and gaining valuable insights into driver behavior. In our study, we use K-Means and MiniSOM as representative Machine Learning (ML) and Deep Learning (DL) algorithms to create 10 clusters from the visited longitudinal and latitudinal locations of all cars in the region, with a specific focus on stops, low speeds and extended staying times features. The choice of these schemes is made as each offers distinct advantages; k-Means efficiently clusters the trajectories as the number of clusters is predefined, while the MiniSOM provides dimensionality reduction and visualization capabilities. The sci-kit-learn Python library is used to implement the k-means. The MiniSOM Python library is used for the SOM model that is trained with an initial learning rate of 0.5, a 0.1 radius of different neighborhoods during training, and 100 epochs; the MiniSom model maps the input data to a 2D 10 $\times$ 1 grid of clusters, and cluster labels are assigned based on their positions in the grid.
Both schemes consider the normalized 'speed,' 'stay,' and the number of visits features. This approach provides interpretable clusters, highlighting patterns like high stay, low speed, and many visits at specific locations, offering valuable insights into the data. After clustering, the mean longitude and latitude coordinates are calculated for each cluster, and each data point is assigned a cluster label. For the found mean coordinates, reverse geocoding retrieves the semantic human-readable address of these locations using the geopy Python library with Nominatim geocoder. The results of each described step in this phase are exhibited using Python's plotly.graph module visualization interface in Fig. \ref{fig:pois1}, Fig. \ref{fig:pois2}, Fig. \ref{fig:pois3}, and Fig. \ref{fig:pois5}. Table \ref{tab:pois} lists the obtained semantic addresses of the extracted POIs from our input dataset.

\bgroup
\renewcommand{\arraystretch}{1}
\begin{table*}[!htbp]
\centering
\caption{POIs Reverse Geocoding Results}\label{tab:pois}
\begin{tabular}{|m{.7cm}|m{1.3cm}|m{1.3cm}|m{13cm}|}\hline
POI&Latitude&Longitude&Address\\\hline
POI 0 POI 6&40.22211679 40.22170348&29.01703473 29.01586844&(3. Yaylı Sokak) / (6. Yeni Sokak), Mehmet Akif Mahallesi, Osmangazi, 16170\\\hline
POI 1 POI 5&40.23167648 40.232187&28.99464074 28.990603&(2. Adalet Caddesi, Adalet Mahallesi) / (Mustafa Münevver Olağaner İlkokulu, Balkan Caddesi, Bağlarbaşı Mahallesi), Osmangazi, 16160\\\hline
POI 2&40.21375&28.99375&Karaman Mahallisi, Nilüfer, 16265\\\hline
POI 3&40.24213565&28.97109287&2. Hürriyet Caddesi, Akpınar Mahallesi, Osmangazi, 16140\\\hline
POI 4 POI 7&40.22732144 40.2263681&29.00227351 29.00469326&(1. Adalet Caddesi) / (Çetin Sokak), Adalet Mahallesi, Osmangazi, 16160\\\hline
POI 8&40.22893693&28.99751062&Hamidiye Mesleki ve Teknik Anadolu Lisesi, 1. Hürriyet Caddesi, Adalet Mahallesi, Osmangazi, 16170\\\hline
POI 9&40.23701047&28.97994714&4. Hakim Sokak, Bağlarbaşı Mahallesi, Osmangazi, 16160\\\hline
\end{tabular}
\end{table*}
\egroup

\subsection{Phase II: VANETs and Their Twins at Crowded POIs}
\begin{figure}[!htbp]
\centering
\includegraphics[width=.6\linewidth]{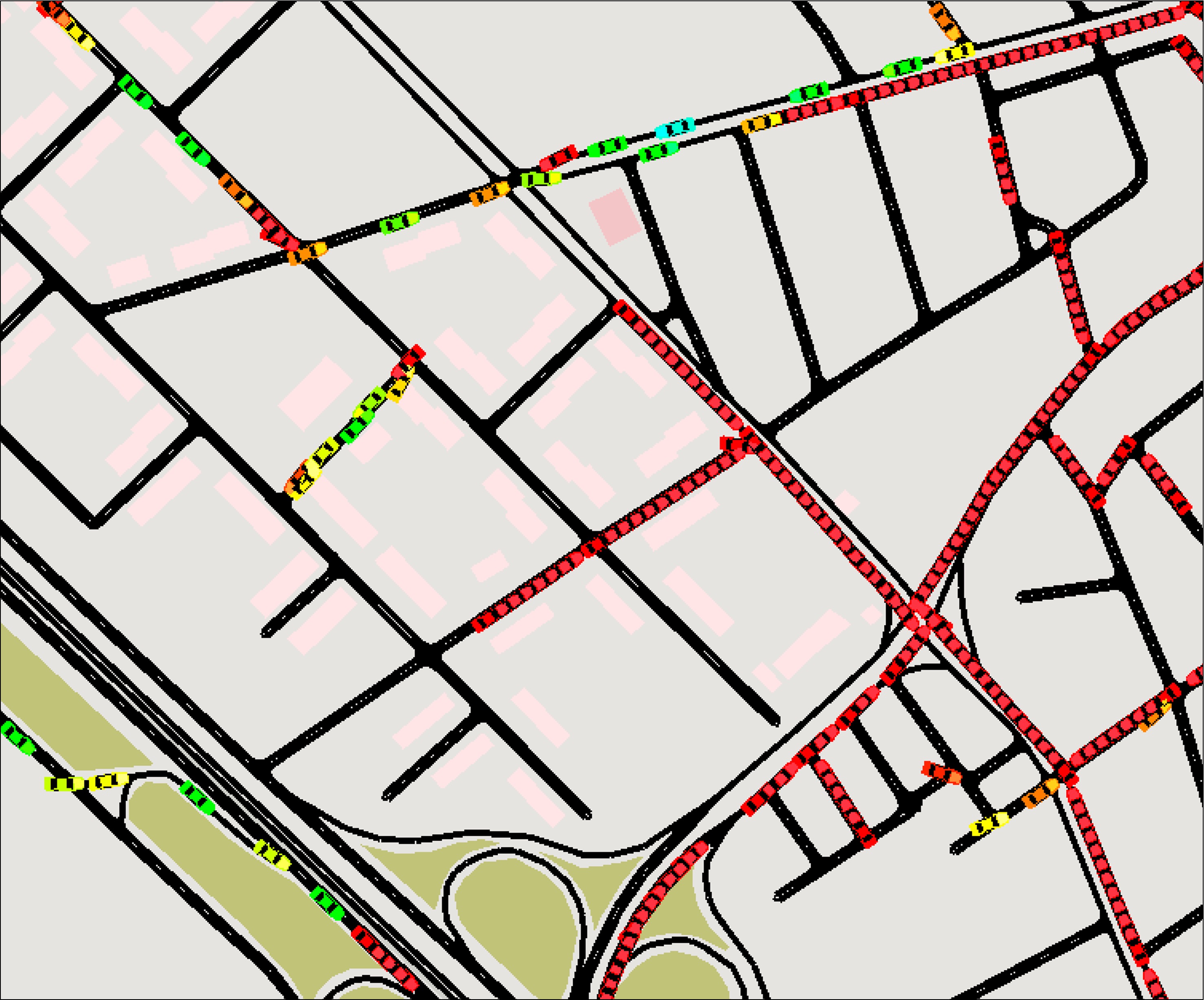}
\caption{Transportation Traffic Status at Extracted POI 3, SUMO Excerpt}
    \label{fig:sumo1}
\end{figure}

Ten POIs are identified in phase I, representing the region's densest and most visited locations. To understand the traffic dynamics, we conducted simulations using the SUMO 1.2.0 environment \cite{lopez2018microscopic}, presenting an instance of transportation traffic at POI 3 in Fig. \ref{fig:sumo1}. However, these crowded areas pose road safety risks and a higher probability of accidents; hence, integrating V2X communication, aka VANETs, with vehicles and infrastructure is encouraged to enhance awareness and decrease witnessed accidents. However, implementing VANETs in dense networks faces the challenges of communication congestion, delays, and data loss when many vehicles contend for communication. Advanced communication protocols, data traffic management strategies, and network optimization techniques are vital to address these challenges. One of the increasingly prominent solutions to address the challenges of crowded environments is creating a virtual DT representation of VANETs \cite{wang2023survey}. Leveraging DTs will offset the load of the physical network to the virtual world server(s); hence, lower communication congestion, delay, and data loss can be witnessed. To support this positional statement, we leverage the POI 3 crowded intersection as a case study for a simple comparative analysis of the physical and virtual networks in cloud, edge, and hybrid deployments. POI 3 covers an area of 0.5 km $\times$ 0.5 km with a single edge Road-Side Unit (RSU), i.e., edge server. The maximum number of vehicles that can connect to the RSU server is limited to 40. For communication, Vehicle to Infrastructure (V2I) alternates between WiFi at a 6Mbps data rate and cellular connections at a data rate of 100 Mbps. The Vehicle to Vehicle (V2V) uses WiFi. The safety messages that the vehicle's beacon every 100 msec require approximately 2.23 msec for signature signing and verification when processed by a regular 2.8 GHz Intel Core i7 CPU. The communicated data size, including the attached cryptographic payload, is 310 bytes. The WiFi has a control channel of 46 msec, on which the vehicles and infrastructure communicate. The reaction latency of the edge server is approximately 0.05 sec, while the cloud server exhibits a reaction latency of 0.1 sec. Through simple, straightforward analysis, we aim to gain valuable insights into the potential advantages of twinning vehicular networks in such dense areas.
\vspace{1em}

\section{Comparative Analysis and Discussions}\label{results}
\begin{figure}[!htbp]
\centering
\includegraphics[width=.3\linewidth]{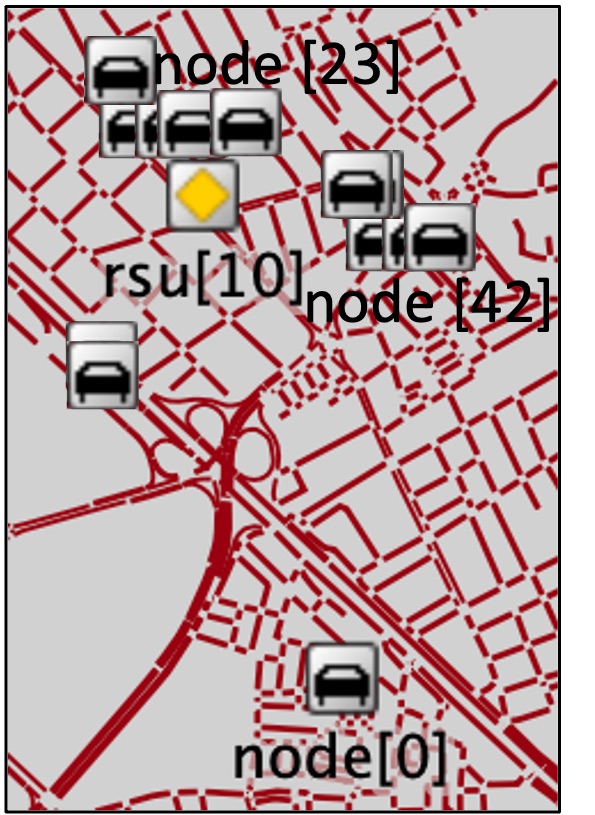}
\caption{OMNET++ POI 3 Simulation Excerpt}
    \label{fig:casestudy}
\end{figure}

This section presents the findings of our comparative study between physical and virtual vehicular networks at the crowded intersection of POI 3. Our case study, Fig. \ref{fig:casestudy}, demonstrates a proof of concept analysis that compares physical VANETs' performance with a cloud-based virtual twin, edge-based vehicular twin, and hybrid-based vehicular twin \cite{luan2021paradigm}. The physical VANET is simulated using the OMNeT++ 6.0.1 simulator \cite{varga2010overview}, utilizing the Bursa map and including all extracted POIs of phase I. Two applications were developed for the taxis (simulation nodes) and POIs (simulation RSUs). The results for POI 3 (RSU 10) are showcased here as an illustrative example. For the virtual VANET twin, we have integrated two servers into our simulations to emulate the edge server for the edge deployment and the cloud server for the cloud deployment. In contrast, the hybrid twin effectively utilizes both servers as required.\vspace{1em}

These different configurations are compared in terms of the following Key Performance Indicators (KPIs):
\begin{itemize}
    \item Latency (sec) measures the time delay experienced by vehicles in the network. It can be defined as the sum of data transmission and processing time at the server(s).
    \item Computation Speed (computations per sec) represents the processing capability of the server(s) regarding the number of computations it can perform per second. It is the reciprocal of the processing time.
\end{itemize}

\begin{figure*}[!htbp]
\centering
\includegraphics[width=.6\linewidth]{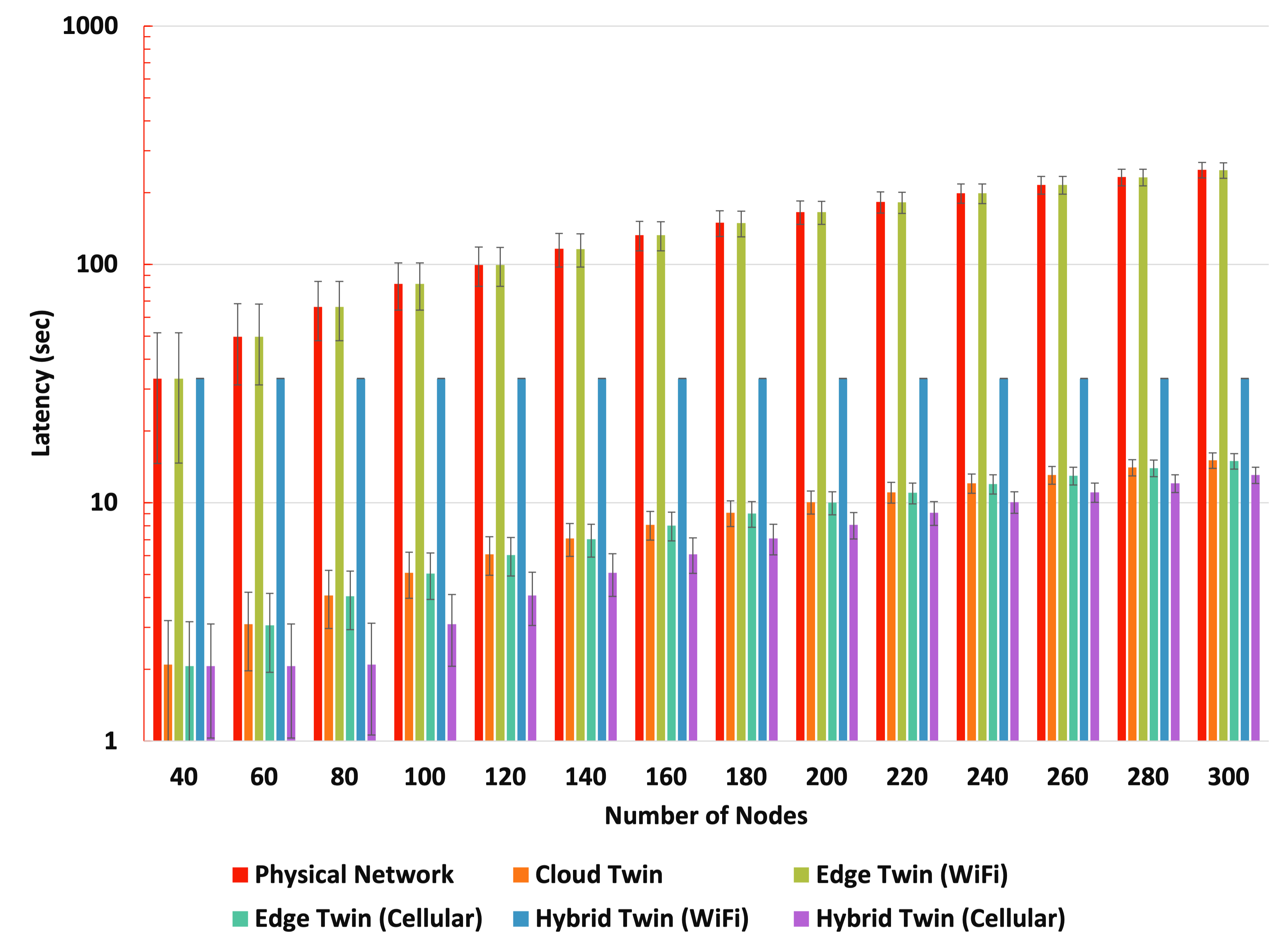}
\caption{Latency (sec) Comparative Analysis}
    \label{fig:latency}
\end{figure*}

The results in Fig. \ref{fig:latency} showcase that twinning can improve the physical network's latency. For example, with varying numbers of simulation nodes at the crowded intersection of POI 3, the physical network demonstrates an expected performance; as the number of vehicles increases, the witnessed latency experiences a significant rise. This behavior is seen as more vehicles contend for the limited communication resources, leading to congestion and increased delays. In contrast, virtual twins show lower delays. For example, the cloud twin and edge twin (Cellular) consistently maintain remarkably low latencies across all vehicle densities, with representative numbers such as 2.0637 sec for 40 vehicles and 15.0469 sec for 300 vehicles. Even as the number of vehicles grows, these twinning approaches efficiently manage latency and ensure smooth communication. Moreover, the edge twin (WiFi), hybrid twin (WiFi), and hybrid twin (Cellular) demonstrate commendable performance, with latencies significantly lower than the physical network while still being slightly higher than the cloud twin and edge twin (Cellular). Latency values of 8.07168 sec for 160 vehicles and 13.05 sec for 300 vehicles highlight the improved communication experience these twins offer compared to the physical network.

On the other hand, a valuable insight into the impact of different twinning strategies on the network's overall computational performance is given in Table \ref{table:speed}. A straightforward analysis of the computation speed shows that in the physical network, computation speed significantly decreases as the number of vehicles increases, ranging from 11.210762 computations per sec for 40 vehicles to 2.2421 computations per sec for 200 vehicles. In contrast, all three deployments of virtual twins consistently exhibit higher computation speeds across all vehicle densities, highlighting the efficient resource management and scalability of virtual twin technologies. Notably, the cloud twin achieves a computation speed of 56.0538, approximately 1.7 times faster than the edge twin's speed of 33.01 computations per second at 80 vehicles. This demonstrates the superior performance of the cloud twin in this particular scenario. However, the hybrid twin strikes a well-balanced performance regardless of network density, making it an attractive option for various scenarios.

\bgroup
\begin{table}[!htbp]
\centering
\renewcommand{\arraystretch}{1}
\caption{Computation Speed (computations per sec) KPI Comparison}
\label{table:speed}
\footnotesize
\begin{tabular}{|m{1.3cm}|m{2cm}|m{1.5cm}|m{2.1cm}|}
\hline
$\#$ Vehicles& Physical Network& Cloud Twin& Edge/Hybrid Twin\\ \hline
40 & 11.210762 & 112.10762 & 33.632287\\\hline
80 & 5.605381166 & 56.05381166 & 33.01\\\hline
120 & 3.736920777 & 37.369207 & 32.2577\\\hline
160 & 2.802690583 & 28.02690583 & 33.632287\\\hline
200 & 2.242152466 & 22.42152466 & 33.632287\\\hline
\end{tabular}
\end{table}
\egroup

Our findings support the superiority of virtual twin deployments for vehicular networks, outperforming traditional physical deployments. These insights provide valuable guidance for implementing twinning strategies that optimize computation and ensure efficient vehicular communication. Embracing virtual twin technologies presents exciting opportunities to revolutionize vehicular networks, especially in crowded places. However, considering real-world environmental factors is essential when making deployment decisions.

\section{Conclusion and Future Work}\label{conc}
This comprehensive study explored vehicular network twins' potential to enhance VANET performance in crowded POIs. Leveraging historical traffic flow data and AI clustering, we identified critical locations requiring efficient vehicular communication for improved road safety. Subsequently, through meticulous simulations and case study analysis, we conclusively demonstrated twinning VANETs as a well-balanced and superior solution, outperforming traditional physical networks. Twin deployments significantly improved network performance with reduced latency and enhanced computation speed. Embracing virtual twin technologies offers exciting possibilities for revolutionizing vehicular networks, especially in dense areas. A future extension will assess vehicular twins' environmental impact, evaluating emissions at crowded locations and AI-driven reduction strategies. This research lays the groundwork for intelligent vehicular network deployment and sustainable urban mobility solutions.
\section*{Acknowledgment}
This work was supported by the ITU Rektorlugu Bilimsel Arastirma Projeleri Birimi (BAP) Fund Grant Number 43981 and by Abu Dhabi National Oil Company (ADNOC), Emirates NBD, Sharjah Electricity Water $\&$ Gas Authority (SEWA), Technology Innovation Institute (TII) and GSK as the sponsors of the $4^{th}$ Forum for Women in Research (QUWA): Sustaining Women's Empowerment in Research $\&$ Innovation at University of Sharjah.
\bibliographystyle{unsrt}
\bibliography{bibe}
\end{document}